# Accounting for gender research performance differences in ranking universities[1]

*Giovanni Abramo (corresponding author)*
  Laboratory for Studies of Research and Technology Transfer
  Institute for System Analysis and Computer Science (IASI-CNR)
  National Research Council of Italy
  Via del Taurini 19, 00185 Rome, Italy
  Tel and fax +39 06 725967362
  giovanni.abramo@uniroma2.it

*Ciriaco Andrea D'Angelo*
  Department of Engineering and Management
  University of Rome "Tor Vergata"
  Via del Politecnico 1, 00133 Rome, Italy
  Tel and fax +39 06 725967362
  dangelo@dii.uniroma2.it

**Abstract**

The literature on the theme of gender differences in research performance indicates a quite evident gap in favor of men over women. Beyond the understanding of the factors that could be at the basis of this phenomenon, it is worthwhile understanding if it would be appropriate to conduct the evaluation per population in a manner distinguished by gender. In fact if there is some factor that structurally determines a penalization of performance by women researchers compared to men then the comparative evaluation of organizations' performance that does not take gender into account will lead to an advantage for those that employ more men, under parity in the capacities of their staffs. In this work we measure the differences of the performance and the rank of research institutions as observed when gender is taken into account compared to when it is ignored. The study population consists of all Italian universities and the performance measured in the hard sciences for the period 2006-2010.

---

**Introduction**

The representation of women in research systems varies across countries, institutions and disciplines. However the fact of their underrepresentation is undeniable. Only four (Portugal, Estonia, Slovak Republic and Iceland) of 28 OECD nations whose data are available show a percentage of women greater than 40%, with a maximum of 46%[1].This phenomenon is underlain by a mix of different factors, with different weights across countries: educational emancipation has come later for women, with consequent lesser numbers of potential candidates for academic positions; lesser interest among women for research activity; the scientific production of women tends to be lower than that for men, perhaps due to the social roles of women as wives and mothers, or from causes of gender discrimination; gender discrimination can also occur in recruitment processes. One way to reduce the underrepresentation of women in the research system is to control for factors exogenous to scientific merit in all processes of comparative performance evaluation, in order to avoid incorrect conclusions and choices that are harmful to women and to institutions with greater female representation in their research staffs. In this regard it is important to note that the so-called "productivity gap" in favor of men is a documented fact. The lesser research performance of females has been established in tens of studies of diverse countries and disciplines[2,3,4,5,6], although it is lessening over time[7,8,9,10,11,12] and it is more visible in the early stages of career[13], and among top scientists[14,15]. Looking at productivity as indicated by patenting, women faculty members produce at about 40% of the rate of men[16].

There is an equally substantial literature investigating the possible causes of the productivity gap, particularly the issues of the environmental and personal factors that can influence the researcher's performance, beyond the personal merit of the individual[17]. Rossiter[18] indicates the particular case of the so-called "Matilda effect", which occurs when female scientists are not recognized in the bylines of the publications resulting from joint research. A separate concern is that in the career recruitment stages the percentages of female applicants who are successful in selection procedures is generally lower[19]. In the subsequent stages of entry to the academic environment females generally evaluate their mentors as less satisfactory than do their male colleagues[20]. However there is also no doubt that the changing personal conditions that the researchers experience over time also affect their productivity. In the late postdoctoral and early faculty years many qualified women scientists stop applying for NIH grants[21]. During their careers, women present lower productivity in the intermediate levels of seniority[22]. In this phase the characteristics of marriage and the presence of school-age children have negative effects on research productivity[23,24,25]. It has been demonstrated that research collaborations have a positive correlation with scientific performance[26] and particularly international collaborations[27], but also that women register less international collaborations than men[28], possibly for reasons of women avoiding longer stays at a distance from their families. Women tend to have more restricted collaboration networks[29,30,31] particularly in the first years of their career[32,33], which limits their access to the resources and assets necessary for their research activity. Duch et al.[34] observe that academic research institutions tend not support women with adequate financial resources, particularly in the hard sciences. According to Ceci and Williams[35] differential gender outcomes result exclusively from differences in resources.

However the aim of the current work is not to investigate further into the issues of if or to what extent there is gender discrimination in the research sphere, or into the objective limitations on women due to their social roles. Rather we wish to determine if the separation of the measurement of research performance by gender produces detectably different results from measurement without gender distinction. In contexts where the potential of discrimination by gender is recognized, or where the family roles of women can condition the time, energy or personal concentration devoted to research, the conduct of comparative evaluation without distinction by gender would inevitably penalize the research organizations that employ a research staff with higher concentration of women. The results of the analysis are thus of certain interest in all processes of comparative evaluation of institutions, such as for example national research assessment exercises, especially where these are intended for the efficient allocation of the available resources. The policy maker can then decide whether the extent of rank differences suggests for gender distinction when conducting institutions' research evaluation exercises. We show evidences for the Italian universities. We proceed by preparing two rankings of the research productivity of the universities for the 2006-2010 period: one obtained through the aggregation of individual performances with distinction by gender and one where the aggregation is conducted without distinction, in order to examine the extent of the differences.

Next section presents the context, the methodology adopted for the calculation of the universities' productivity and the dataset used for the analyses. A section setting out the main results of the work follows. The final section presents the conclusions and offers several policy indications.

**Context, method, and data**

The Italian Ministry of Education, Universities and Research (MIUR) recognizes a total of 96 universities as having the authority to issue legally recognized degrees. Of these, 29 are small, private, special-focus universities, of which 13 offer only e-learning; 67 are public and generally multi-disciplinary universities, scattered throughout Italy. In keeping with the Humboldtian model, there are no 'teaching-only' universities in Italy, as all professors are required to carry out both research and teaching. In the Italian university system all professors are classified in one and only one field (named the scientific disciplinary sector, SDS, 370 in all). Fields are grouped into disciplines (named university disciplinary areas, UDAs, 14 in all). The overall staff system of over 58,000 professors, of which 95% are employed in public universities. The makeup of faculty members features a majority of men, although the data since 1998 indicate a trend towards increasing presence of women. This shows also in the representation of female assistant professors (45.3%), which is now much higher than that of full professors (20.7%). Female professors are in the majority only in the UDAs of Ancient history, philology, literature, art history (55.2%) and Biology (51.6%). The UDAs with the lowest presence of women are Physics (19.6%) and Industrial and information engineering (15.1%).

*Measuring research productivity of universities*

To measure research productivity of universities we adopt the approach thoroughly described in Abramo and D'Angelo[36]. We begin by measuring research productivity at the individual level and then aggregate the individual measures for the evaluation of organizations. At the individual level, we adopt an indicator named Fractional Scientific Strength (FSS) embedding both the number of publications produced, their standardized impact and number of co-authors of each one. In formula, the average yearly productivity of an individual, over a period of time, accounting for the cost of labor, is:

$$FSS = \frac{1}{w} * \frac{1}{t} \sum_{i=1}^{N} \frac{c_i}{\bar{c}} f_i \qquad [1]$$

Where:
w = average yearly wage of the professor;
t = number of years of work by the professor in the period under observation;
N = number of publications by the professor in the period under observation;
$c_i$ = citations received by publication *i*;
$\bar{c}$ = average of the distribution of citations received for all cited publications indexed in the same year and subject category of publication *i*;
$f_i$ = fractional contribution of the researcher to publication *i*. The fractional contribution equals the inverse of the number of authors, except that in life sciences, where the various contributions are weighted according to the order of the names in the byline[37].

Research productivity of universities is obtained by aggregating individual research productivity, according to the following formula:

$$FSS_U = \frac{1}{RS} \sum_{i=1}^{RS} \frac{FSS_{R_i}}{\overline{FSS_{R_i}}} \qquad [2]$$

Where:
$RS$ = research staff of the university, in the observed period;
$FSS_{R_i}$ = productivity of researcher *i* in the university;
$\overline{FSS_{R_i}}$ = average productivity of all productive researchers in the same SDS of researcher *i*.

The choice of average productivity of all productive researchers as the optimum scaling factor to reduce distortions when comparing performance of heterogeneous research institutions is based on the results of a study by Abramo et al.[38]. This scaling factor could be calculated separately for the two subpopulations of genders, since Abramo et al.[39] have demonstrated that the relative distributions result as significantly different. We ask what are the effects on the value of $FSS_U$ (and thus on the ranking lists) from the choice of whether or not to apply a scaling factor differentiated by gender. We will attempt to provide an answer to this question in the "Analysis and discussion" section, however first we illustrate the dataset used in the analyses.

*Data and sources*

Data on the research staff of each university, such as years of employment in the observed period, academic rank and their SDS classification, are extracted from the database on Italian university personnel maintained by the Ministry for Universities and Research. Unfortunately, information on leaves of absence is not available and cannot be accounted for in the calculation of yearly productivity, to the disadvantage of women on maternity leave in the period of observation.

The bibliometric dataset for the analysis draws on the Observatory of Public

Research (ORP), a database developed and maintained by the authors and derived under license from the WoS. Beginning from the raw data of Italian publications indexed in WoS, and applying a complex algorithm for disambiguation of the true identity of the authors and their institutional affiliations (for details see D'Angelo et al., 2011), each publication is attributed to the university professor that produced it, with a harmonic average of precision and recall (F-measure) equal to 96 (error of 4%). Beginning from this data we are able to calculate FSS for each Italian professor. We limit the field of observation to the hard sciences, i.e. the first 9 UDAs in Table 1. For the WoS-indexed publications to serve as a more robust proxy of overall output of a researcher, the field of observation is further limited to those SDSs (188 in all) where at least 50% of member scientists produced at least one publication in the period 2006-2010. For the purposes of the study and to ensure significant representation of both genders in each field, we then limit the analysis to those SDSs (99 in all) with at least 30 individuals of each gender. Finally, for a robust comparison of university ranks by UDA, we exclude those universities with less than 10 professors in the UDA. Table 1 shows the final dataset.

*Table 1: Dataset of the analysis: number of fields (SDSs), universities and professors in each UDA under investigation*

| UDA | N. of SDSs | Professors | | Universities | |
|---|---|---|---|---|---|
| | | Total | Female | Total | With at least 10 professors |
| Mathematics and computer science | 8 | 3,297 | 1,105 (33.5%) | 65 | 50 |
| Physics | 4 | 2,161 | 390 (18.0%) | 61 | 43 |
| Chemistry | 9 | 3,199 | 1,212 (37.9%) | 59 | 41 |
| Earth sciences | 4 | 534 | 176 (33.0%) | 41 | 22 |
| Biology | 19 | 5,338 | 2,591 (48.5%) | 66 | 50 |
| Medicine | 29 | 9,426 | 2,805 (29.8%) | 60 | 42 |
| Agricultural and veterinary sciences | 17 | 2,163 | 755 (34.9%) | 43 | 27 |
| Civil engineering | 3 | 828 | 130 (15.7%) | 49 | 31 |
| Industrial and information engineering | 6 | 2,051 | 298 (14.5%) | 64 | 42 |
| Total | 99 | 28,997 | 9,462 (32.6%) | 79 | 64 |

**Analysis and discussion**

As an example of the preparation of the ranking lists, Table 2 shows that for the Italian universities active in chemistry, ranked for productivity. As noted, for reasons of significance the construction of the ranking list considers only those universities (41 in Chemistry) with at least 10 professors in the UDA. For each university, the table shows: the absolute values of productivity calculated as in [2] with and without gender distinction, the relative positions in the ranking and the differences that emerge in terms of value and sign. Eight of the universities listed maintain the same position in the ranking, however 33 show changes. Sixteen of the 33 move up in the ranking taking account of gender and among these, two (Univ_13 and Univ_18) gain four positions, while another two (Univ_5 and 21) gain three places. On the opposite side we find three universities that lose four positions (Univ_3, 15 and 22) and two that lose three places (Univ_14 and 19).

Considering that the value of FSS for the UDA is given by the average of the individual values rescaled to the average of their SDS, we can apply the t-test for paired samples in each university to evaluate the significance of any differences between the values of productivity. In formula, the t-test applied is:

$$t-test_{paired} = \frac{FSS_U^1 - FSS_U^2}{s/\sqrt{n}}$$

[3]

Where:
$FSS_U^1$ = university's productivity without gender distinction
$FSS_U^2$ = university's productivity with gender distinction
$s$ = standard deviation of the difference between $FSS_U^1$ and $FSS_U^2$
$n$ = number of researchers present in the UDA

Table 2 presents the results of this test, with the asterisks in columns 1 and 8 indicating the universities that show significant differences in the two rankings. In the Chemistry area, 11 universities out of 41 show significant differences in productivity when distinguished for gender, with these differences having only very partial effect on the variation in rank, as demonstrated by the high value of Spearman correlation (0.986) and the low average number of variations (1.561).

*Table 2: Productivity rankings of Italian universities in Chemistry (2006-2010) with ($FSS_U^2$) and without ($FSS_U^1$) gender distinction*

| Univ. | $FSS_U^1$ Abs. value | Rank | $FSS_U^2$ Abs. value | Rank | Sign | rank diff. (abs) | Univ. | $FSS_U^1$ Abs. value | Rank | $FSS_U^2$ Abs. value | Rank | Sign | rank diff. (abs) |
|---|---|---|---|---|---|---|---|---|---|---|---|---|---|
| 1 | 1.724 | 1 | 1.737 | 1 | = | 0 | 22*** | 0.805 | 22 | 0.750 | 26 | - | 4 |
| 2 | 1.307 | 2 | 1.320 | 3 | - | 1 | 23** | 0.800 | 23 | 0.842 | 21 | + | 2 |
| 3** | 1.289 | 3 | 1.168 | 7 | - | 4 | 24 | 0.771 | 24 | 0.801 | 23 | + | 1 |
| 4 | 1.283 | 4 | 1.287 | 5 | - | 1 | 25 | 0.769 | 25 | 0.794 | 24 | + | 1 |
| 5 | 1.256 | 5 | 1.325 | 2 | + | 3 | 26* | 0.745 | 26 | 0.783 | 25 | + | 1 |
| 6* | 1.196 | 6 | 1.293 | 4 | + | 2 | 27 | 0.718 | 27 | 0.732 | 27 | = | 0 |
| 7 | 1.190 | 7 | 1.160 | 8 | - | 1 | 28 | 0.712 | 28 | 0.715 | 28 | = | 0 |
| 8 | 1.174 | 8 | 1.180 | 6 | + | 2 | 29 | 0.701 | 29 | 0.667 | 30 | - | 1 |
| 9 | 1.121 | 9 | 1.107 | 10 | - | 1 | 30 | 0.680 | 30 | 0.660 | 31 | - | 1 |
| 10 | 1.091 | 10 | 1.042 | 12 | - | 2 | 31 | 0.661 | 31 | 0.671 | 29 | + | 2 |
| 11 | 1.076 | 11 | 1.039 | 13 | - | 2 | 32* | 0.642 | 32 | 0.619 | 33 | - | 1 |
| 12 | 1.056 | 12 | 1.094 | 11 | + | 1 | 33 | 0.634 | 33 | 0.622 | 32 | + | 1 |
| 13 | 1.024 | 13 | 1.112 | 9 | + | 4 | 34 | 0.627 | 34 | 0.614 | 34 | = | 0 |
| 14* | 0.988 | 14 | 0.918 | 17 | - | 3 | 35*** | 0.613 | 35 | 0.584 | 35 | = | 0 |
| 15 | 0.985 | 15 | 0.877 | 19 | - | 4 | 36 | 0.565 | 36 | 0.527 | 37 | - | 1 |
| 16 | 0.984 | 16 | 0.938 | 16 | = | 0 | 37 | 0.540 | 37 | 0.496 | 38 | - | 1 |
| 17 | 0.976 | 17 | 0.956 | 15 | + | 2 | 38 | 0.540 | 38 | 0.553 | 36 | + | 2 |
| 18 | 0.953 | 18 | 0.991 | 14 | + | 4 | 39 | 0.489 | 39 | 0.431 | 40 | - | 1 |
| 19** | 0.928 | 19 | 0.835 | 22 | - | 3 | 40 | 0.459 | 40 | 0.487 | 39 | + | 1 |
| 20 | 0.882 | 20 | 0.863 | 20 | = | 0 | 41 | 0.255 | 41 | 0.261 | 41 | = | 0 |
| 21 | 0.862 | 21 | 0.890 | 18 | + | 3 | | | | | | | |

N° of observations: 41; Sum of differences: 62; Max of difference: 4; Mean of differences: 1.561.
*** paired t test: p-value<0.01; ** p-value<0.05; * p-value<0.10

The above analyses are repeated for each of the remaining eight UDAs, with Table 3 showing the summary results. In comparing the rankings it emerges that Biology has the highest percentage of universities that change at least one position (84%), while Physics has the lowest percentage (44%). We recall from "Data and sources" section that physics is also the UDA with the least representation of women. Mathematics and computer sciences shows the cases of the highest individual shifts, with one university gaining nine positions in the productivity ranking by gender and another that loses nine positions. This UDA also registers the highest average shift per university (2.2 positions). This UDA contrasts to Agricultural and veterinary sciences, where the maximum shift in position is 2 and the average shift is 0.519. The quite limited variations in rank are accompanied by Spearman correlation values that are consistently above a 0.96, and all highly significant. Applying the *t-test* we observe another notable result: the highest number of universities with significant differences in productivity with and without gender distinction is seen in both Chemistry (11 out of 41) and in Physics (11 out of 43), however in the latter UDA the differences have minimal impact on the rankings.

*Table 3: Descriptive statistics of rank differences between $FSS_U$, calculated with and without gender distinction*

| UDAs | N° of universities‡ | Shifting in rank | Max shift | Average shift | Spearman ρ |
|---|---|---|---|---|---|
| Agricultural and veterinary sciences | 27 (2) | 12 (44.4%) | 2 | 0.519 | 0.995*** |
| Biology | 50 (7) | 42 (84.0%) | 7 | 1.960 | 0.984*** |
| Chemistry | 41 (11) | 33 (80.5%) | 4 | 1.561 | 0.986*** |
| Civil engineering | 31 (8) | 18 (58.1%) | 3 | 0.710 | 0.994*** |
| Earth sciences | 22 (5) | 15 (68.2%) | 5 | 1.091 | 0.968*** |
| Industrial and information engineering | 42 (8) | 21 (50.0%) | 4 | 0.857 | 0.994*** |
| Mathematics and computer sciences | 50 (10) | 39 (78.0%) | 9 | 2.200 | 0.978*** |
| Medicine | 42 (5) | 20 (47.6%) | 5 | 1.000 | 0.990*** |
| Physics | 43 (11) | 19 (44.2%) | 3 | 0.605 | 0.997*** |

*Significance level: \*\*\*= p < 0.01; \*\*=p < 0.05; \*= p<0.10.*
‡in brackets the number of universities with significant differences (p-value<0.10) in $FSS_U$ with and without gender distinction

As an alternative to the Spearman correlation coefficient, which measures the intensity and sign of the interdependence between the two ranking lists, we also consider another indicator ($R'$) that measures the potential of the rank differences. This is given by the sum of the absolute differences in rank registered in an area and the maximum sum of rank differences in reference to the theoretical situation of perfect inversion of the rankings. The indicator assumes nil value in case of identical ranking lists with and without gender distinction and 100 in the case of perfect inversion of the ranking lists. In formula:

$$R' = \frac{\sum_{i=1}^{n}|d_{rank_i}|}{max}$$

[4]

Where:
$d_{rank_i}$ = difference in rank registered for university *i*, under the two evaluation methods
$n$ = number of universities active in the UDA.

$$max = \begin{cases} \frac{n^2}{2}, \text{for } n \text{ even number} \\ (n-1) * \left(\frac{n-1}{2} + 1\right), \text{for } n \text{ odd number} \end{cases}$$

Table 4 presents the example of a fictitious case of five universities: comparing two hypothetical rankings (Rank#1 and Rank#2), we obtain the sum of the absolute differences in rank as being 6. In the case of perfect inversion of the rankings (Rank#1$_{inverted}$) the sum of the differences would be 12, from which we obtain an $R'$ value (ratio of 6 to 12) equal to 50%.

Figure 1 presents the $R'$ values for each UDA: in no case does the indicator exceed 10%. In Physics the differences in rank are the lowest in comparison with the other UDAs ($R'$=2.8%). This UDA and another, Agricultural and veterinary sciences ($R'$=3.8%), form a first cluster with quite low values for shifts in ranking. A second group of UDAs with higher values of $R'$ but still lower than 5% is composed of Industrial and information engineering, Civil engineering, and Medicine. A third cluster, with values between 7% and 8% is composed of Biology and Chemistry. A final cluster, composed of Mathematics and Earth sciences, shows values of the indicator over 8%.

*Table 4: Example of calculation of the indicator R'*

| University | Rank#1 | Rank#2 | Abs. Diff. \|Rank#1 - Rank#2\| | Rank#1$_{inverted}$ | Abs. Diff. \|Rank#1 - Rank#1$_{inverted}$\| |
|---|---|---|---|---|---|
| ID1 | 1 | 2 | 1 | 5 | 4 |
| ID2 | 2 | 3 | 1 | 4 | 2 |
| ID3 | 3 | 4 | 1 | 3 | 0 |
| ID4 | 4 | 1 | 3 | 2 | 2 |
| ID5 | 5 | 5 | 0 | 1 | 4 |
| | Total differences | | 6 | | 12 => R'= 6/12 = 50% |

*Figure 1: Values (%) of the indicator R' per UDA*

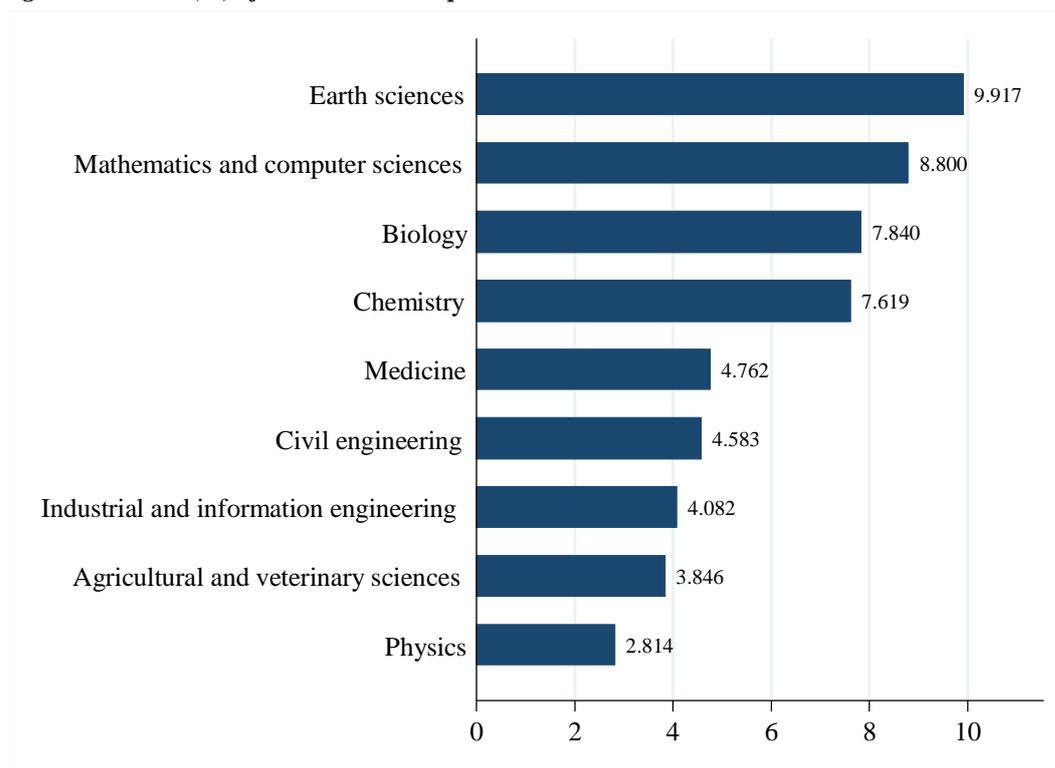

- Earth sciences: 9.917
- Mathematics and computer sciences: 8.800
- Biology: 7.840
- Chemistry: 7.619
- Medicine: 4.762
- Civil engineering: 4.583
- Industrial and information engineering: 4.082
- Agricultural and veterinary sciences: 3.846
- Physics: 2.814

**Conclusions**

Higher education institutions represent an important pillar of national research and innovation systems, thus the policy agendas of many countries now place high priority on strengthening such institutions. One expression of this policy is the increasing diffusion of national research assessment exercises. Such assessments serve different goals, including a strategic one related to efficient resource allocation and stimulation of performance improvement. It follows that they must be conceived and executed with maximum methodological rigor.

Stimulated by a now well consolidated literature that indicates the presence of a "productivity gap" in favor of male researchers, in this work we have attempted to verify the impact of gender aspects in the outcomes of bibliometric assessments carried out for research institutions.

In theory in fact, if there is some factor that structurally determines a penalization of performance for women researchers compared to men, then a comparative evaluation of organizational performance that does not take gender into account will lead to an advantage for those that employ more men, under parity of capacities in the research staffs.

The analyses conducted concerned Italian universities active in the hard sciences for the period 2006-2010. Differently from our previous study on individual scientists rankings[39], the results here show a very strong correlation (never below 0.96) between the two ranking lists: the one that did not distinguish by gender and the one that did. As could be expected, the gender productivity gap tends to have limited impact on the comparative aggregated performance of an entire organization, in part certainly because at the level of entire disciplinary areas the distribution of genders among the universities is not particularly heterogeneous.

Still, we should not ignore some of the shifts in performance observed at the level of the specific university disciplinary areas. For example, in Mathematics and computer sciences, 10 of the 50 universities evaluated show significantly different productivities under the two methods of evaluation, and two of these universities shift a full nine positions. The shifts in positions in Biology and Earth sciences are not at all negligible, while the generally high levels of correlation between the rankings also hide diverse and important outliers.

If the objectives of the national evaluation exercises are to stimulate improvement in the general performance in the system, to permit the users to make informed choices, and eventually to guide the allocation of resources (as in Italy and in a growing number of other countries), it is important that all factors exogenous to the true merit of the subjects evaluated be held in due account and appropriately controlled for. This does not mean that we intend to issue *a priori* recommendations on the suitability of conducting comparative evaluation research performance that would take account of gender. Our current objective was to measure to what extent the comparison of institutions' research performance with and without distinguishing by gender leads to rank positions that are detectably different We leave it to the decision-maker to choose which approach to adopt, given the objectives of the evaluation and the conditions of the context. He/she should also consider that the choice to conduct evaluation exercises distinguished by gender may itself be interpreted by female scientists as a form of unnecessary and unwanted discrimination. Female researchers might in fact perceive the procedure as implying a cognitive gap in favor of men.

**Acknowledgements**

The work has received no financial support by third parties.

**Conflict of interest**

The authors declare no conflict of interest.